\documentclass[10pt,twocolumn,letterpaper]{article}

\usepackage{iccv}
\usepackage{times}
\usepackage{epsfig}
\usepackage{graphicx}
\usepackage{amsmath}
\usepackage{amssymb}

\usepackage[linesnumbered, vlined, ruled, commentsnumbered]{algorithm2e}

\usepackage{multirow} 
\usepackage{xcolor}
\usepackage{booktabs}
\usepackage{subcaption}
\usepackage{array}
\newcolumntype{C}[1]{>{\centering\arraybackslash}p{#1}}

\usepackage{stackengine}
\newcommand\xrowht[2][0]{\addstackgap[.5\dimexpr#2\relax]{\vphantom{#1}}}

\newcommand{\ctr}[1]{\textcolor{red}{#1}}
\newcommand{\ctb}[1]{\textcolor{blue}{#1}}

\usepackage[pagebackref=true,breaklinks=true,letterpaper=true,colorlinks,bookmarks=false]{hyperref}

\iccvfinalcopy 

\ificcvfinal\fi

\newcommand{\skipspace}{\vspace{0pt}}
\begin{document}

\title{Rethinking Noise Synthesis and Modeling in Raw Denoising}
\author{
Yi Zhang\textsuperscript{1}
\and Hongwei Qin\textsuperscript{2}
\and Xiaogang Wang\textsuperscript{1}
\and Hongsheng Li\textsuperscript{1} \smallskip
\and
\textsuperscript{1}CUHK-SenseTime Joint Lab\qquad 
\textsuperscript{2}SenseTime Research \\
{\tt\small zhangyi@link.cuhk.edu.hk,  qinhongwei@sensetime.com, \{xgwang, hsli\}@ee.cuhk.edu.hk}
}

\maketitle
\ificcvfinal\thispagestyle{empty}\fi

    \begin{abstract}
        The lack of large-scale real raw image denoising dataset gives rise to challenges on synthesizing realistic raw image noise for training denoising models.
        However, the real raw image noise is contributed by many noise sources and varies greatly among different sensors.
        Existing methods are unable to model all noise sources accurately, and building a noise model for each sensor is also laborious.
        In this paper, we introduce a new perspective to synthesize noise by directly sampling from the sensor's real noise.
        It inherently generates accurate raw image noise for different camera sensors.
        Two efficient and generic techniques: pattern-aligned patch sampling and high-bit reconstruction help accurate synthesis of spatial-correlated noise and  high-bit noise respectively.
        We conduct systematic experiments on SIDD and ELD datasets. The results show that  (1) our method outperforms existing methods and demonstrates wide generalization on different sensors and lighting conditions. (2) Recent conclusions derived from DNN-based noise modeling methods are actually based on inaccurate noise parameters. The DNN-based methods still cannot outperform physics-based statistical methods.  The code will be available at \url{https://github.com/zhangyi-3/noise-synthesis}.
    \end{abstract}

    \section{Introduction}
    Learning-based methods have made great progress on raw image denoising in recent years. However, collecting a large-scale real raw image dataset is time and manpower consuming.
    Therefore, most learning-based methods on raw image denoising are trained with synthetic datasets, which model and apply synthesis noise to clean images to create noisy inputs. As a result, their denoising performances in real-world scenarios are significantly dependent on the discrepancy between the synthetic noise and actual noise in real raw images.

    Existing methods for synthesizing raw image noise generally conducts the following two steps: (1) building a noise model and optimizing the parameters by fitting the real noise distribution, and (2) generating the synthetic noise randomly from the noise model.
    According to the different types of noise models, they can be divided into two categories: physics-based statistical methods and Deep Neural Network (DNN)-based methods.

    DNN-based methods \cite{dualNoiseGan20, noiseflow,CameraNoiseModeling} learn to model noise distribution from real datasets with deep generative networks (\eg, GAN, Flow). Although the deep models have powerful representation capability, it is particularly hard for them to generate accurate values for each pixel in dense-prediction tasks. For example, in image generation, how to generate realistic small details with less artifacts is a relatively challenging problem.
    While all DNN-based methods claim superior performances on SIDD dataset compared with the physics-based noise models, we found that those conclusions are built on the uncalibrated noise profiles of the SIDD dataset (noise profiles for the Poisson-Gaussian distribution).

    Physics-based statistical methods \cite{eld, FoiPG} are more promising compared with the DNN-based methods since
    they follow the physical process of the specific camera sensor and model various noise sources step by step according to the physical process from photos to digital numbers.
    However, there also exist obvious limitations for physics-based methods. First, it is impossible to accurately extract and model all kinds of noise sources in the camera since every electronic component of the camera can be a noise source.
    Second, in most cases, modeling a noise source is based on inaccurate distributions since the natural distributions for most noise sources are unknown.
    Third, the real noise distributions vary dramatically in different cameras or  lighting conditions (\eg, low-light), which makes physics-based methods laborious. All those limitations hinder physics-based methods to achieve accurate noise modeling.

    In this paper, we propose a new perspective for synthesizing realistic image noise. Specifically, instead of building the noise model first and then generating synthetic noise from the noise model step by step, we synthesize the noise by sampling directly from the real noise distribution.

    We first describe the general raw image formation based on the physical process and decompose the raw image noise into signal-dependent and signal-independent components.
    For the signal-dependent noise, we only consider the photon shot noise, which follows the Poisson distribution strictly due to the quantum nature of light. Other signal-dependent noise (\eg, Photo Response Non-Uniformity) only have marginal impact\footnote{Several previous papers and experiments show the percentage of the mean signal value of PRNU is mostly less than 3\% (\eg Figure~2 in \cite{gow2007comprehensive} and Figure~7 in \cite{prnu2}.) } due to the advances of sensor design and manufacture. We synthesize the signal-dependent noise by sampling from real Poisson distribution with the calibrated total gain parameters.
    For the complicated signal-independent noise, we use the target sensor to capture dark frames with the corresponding exposure and ISO settings in a purely dark room so that all signal-independent noise are preserved and saved into a dark-frame database. We synthesize the signal-independent noise by directly sampling pixel values from the dark-frame database.

    Even though the naive implementation that samples pixel values from 10-bit dark frames is based on the real noise, we empirically find that it cannot work well over all camera sensors and exposure ratios due to the missing of spatial-correlated noise and high-bit information.
    We therefore propose two techniques to tackle the issues: First, we use pattern-aligned patch sampling for signal-independent noise to keep the spatial-correlated and pattern-correlated noise (\eg, row noise, fixed pattern noise).
    Second, we reconstruct the accurate continuous distribution for the quantized dark frames by using a set of bell-shaped distributions and restore the signal-independent noise to higher bits.

    To demonstrate the effectiveness and generalization of our method, we systematically compare it with both DNN-based and statistical methods on SIDD \cite{sidd} and SID \cite{sid} datasets. Our method outperforms all existing methods and shows consistent improvements on various camera sensors and lighting conditions.

    Moreover, we find that the noise profiles of popular raw image datasets \cite{sidd, sid} are quite inaccurate, which have been used widely in previous works.
    After calibrating the noise profile carefully for the statistical methods,
    we observe that recent DNN-based methods actually underperform statistical methods with a clear gap, which is opposite to recent works.

    Our contributions are summarized as follows:
    \begin{itemize}
        \item We propose an efficient and generic perspective to synthesize noise for raw image denoising.

        \item We propose raw pattern-aligned patch and high-bit reconstruction to synthesize realistic raw image noise. 

        \item We systematically study the statistical methods and DNN-based methods for noise modeling.

        \item The proposed method outperforms existing statistical and DNN-based methods on raw image datasets and demonstrates great generalization on different camera sensors and lighting conditions.
    \end{itemize}

    \section{Related Work} 
    Image denoising has been studied for many years.
    In this section, we only review the related works for image noise modeling. The methods have been categorized into two types: physical-based and DNN-based.
    \subsection{Physics-based statistical noise modeling}
    The Gaussian noise model is most widely used when we evaluate the denoising methods \cite{dncnn, n2n, bm3d, bm3d2019, nlm}. But its generalization in real-world denoising is relatively poor.
    For real-world raw image noise, the Poisson-Gaussian (P-G) distribution \cite{FoiPG, FoiClipPG} is one of the typical noise models. It models the shot and read noise by Poisson and Gaussian distribution. A widely used alternative of P-G distribution is to approximate the Poisson distribution to the Gaussian distribution \cite{unprocessing, kpn}. The overall noise model turns to a heteroscedastic Gaussian model with zero mean and signal-dependent variance. But recent work \cite{eld} shows that using the accurate P-G noise model improves the denoising results over using the heteroscedastic Gaussian model.

    The noise modeling for extreme low-light environments is also important for low-light raw image and video denoising. Recent works \cite{enhancingNJU, eld} build the noise model for the low-light condition by analyzing the sensor processing pipeline and modeling noise sources with the assumed distributions.
    But, as studied in the electronic imaging community \cite{HealeyK94, KonnikW14, gow2007comprehensive, el2005cmos}, the camera sensor processing pipeline is relatively complicated, which makes it impossible to extract and model all noise sources.
    
    \subsection{DNN-based noise modeling}
    Another research direction focuses using Deep Neural Network (DNN) to model the real noise distribution.
    Early methods \cite{cvpr18BlindGan, GRDN} use GAN to generate the noise distribution but only show limited performance improvements. More recent works demonstrate their effectiveness on the SIDD dataset \cite{sidd}.
    Noise FLow \cite{noiseflow} following the sensor processing pipeline adopts a flow-based generative model to generate raw image noise. Camera-aware GAN \cite{CameraNoiseModeling} learns a camera-aware model that regards the synthetic noisy images by P-G noise model as the generator input. All of them claim superior performance compared with the physics-based methods. But we found the in-camera noise profiles for the physics-based methods are inaccurate. After careful calibration, the physics-based methods outperform the DNN-based methods.

    \section{Method}

    \subsection{Raw Image Formation}

    To analyze the noise sources in raw images, we first describe the raw image formation of the Complementary Metal-Oxide-Semiconductor (CMOS) sensors, which is the dominant sensor of both Digital Single Lens Reflex Cameras (DSLRs) and smartphones.
    For a raw image produced by a typical CMOS sensor, we model
    the process\footnote{Since the saturation effect can be implemented exactly by the clipping operation, we didn't include it in our noise modeling.} from incident photons to digital values as
    \begin{equation}
        \begin{aligned}
            D = (K_a (I + N_p + N_1) + N_2 + N_q) K_d, \label{eq:overview_eq}
        \end{aligned}
    \end{equation}
    where $D$ is the digital values saved in raw images, $N_q$ is the corresponding quantization noise when transforming the analog signals to digital values, $I$ is the number of incident photons of the real scene, and $N_p$ is the signal-dependent photon shot noise.
    $K_a$ and $K_d$ are analog gain and digital gain respectively. $N_1$ and $N_2$ denote the summation of the other noises produced before the analog gain and digital gain.
    We will then explain all noise in Eq.~\eqref{eq:overview_eq} in detail.

    Due to the the quantum nature of light and the uncertainty of the collected photons, the number of incident photon numbers $(N_p + I)$ for all pixels follow the Poisson distribution strictly:
    \begin{equation}
        (N_p + I) \sim \mathcal{P}(I) \label{eq:shot_noise},
    \end{equation}
    where $\mathcal{P}$ denotes the Poisson distribution that regards the photon number of real scenes $I$ as the expectation.

    Before saving the raw signal to fixed-bit raw images (\eg, 10- or 14-bit), the Analog-to-Digital Converters (ADC) quantizes the analog signals with the fixed quantization step:
    \begin{equation}
        \begin{aligned}
            &D(x, y) = nq, \quad 0 \leq n \leq 2^b - 1, \\
            &N_q \sim \mathcal{U}(-\frac{1}{2} q,\frac{1}{2} q),
        \end{aligned}
    \end{equation}
    where $D(x, y)$ is the value in the position $(x, y)$, $b$ is the number of bits, $q$ is the quantization step, $n$ is the quantization step number, $N_q$ is the quantization noise following the uniform distribution $\mathcal{U}(-\frac{1}{2}q, \frac{1}{2}q)$.

    Since the gain factor in the physical processes affects the noise distribution significantly, we divide the sensor processing into stages according to the position of gain factors. $N_1$ and $N_2$ represent the summation of the other noise before the analog gain and the digital gain respectively. Their components are complicated and vary greatly on different sensors.
    Generally, for the noise $N_1$, it includes the noise introduced by the dark current (dark current shot noise, and fixed pattern noise) \cite{baer2006model}, reset noise \cite{KonnikW14}, flicker noise \cite{barnes1966statistical}, etc.
    For the noise $N_2$, it includes thermal noise, column fixed pattern noise \cite{snoeij2006cmos}, and so on.

    \subsection{Noise synthesis for signal-dependent and signal-independent noise}
    Through the above discussed raw image formation, we can see that the noise distribution in raw images is quite complicated and highly relies on the sensor processing pipeline.
    As a result, the difficulty of building the noise model lies in two aspects. First, since every semiconductor and operation in the circuit can be a noise source and we do not know what is the most appropriate distribution for most of them, it is impossible to isolate and model each noise accurately for all noise sources.
    Second, different sensor processing pipelines and different lighting conditions also affect the actual noise distribution significantly. Building and calibrating the noise model for each case is laborious, especially when we aim to synthesize accurate noise for a wide range of devices and lighting conditions.

     Our key insight is that we can decompose the raw images' noise into signal-dependent noise and signal-independent noise. Signal-dependent noise can be mostly modeled as Poisson distribution and its synthesis can be easily achieved by sampling from the distribution. Signal-independent noise can be synthesized via sampling from an actual noise database rather than building a mathematical noise model and synthesizing noise from the model.
    In this way, no matter what the sensor processing pipeline and lighting conditions are, the noise can be synthesized much more accurately.
 
    Specifically, we analyze the image formation and decompose the raw image noise in Eq.~\eqref{eq:overview_eq} into signal-dependent and signal-independent components:
    \begin{equation}
        D = \underbrace{K_d K_a (I + N_p)}_{\text{signal-dependent}} + \underbrace{K_d K_a N_1 + K_d N_2 + K_d N_q}_{\text{signal-independent}}.
        \label{eq:decomposed_overview}
    \end{equation}
    \paragraph{Signal-dependent noise.} For the signal-dependent noise, we only consider the photo shot noise in the raw image formation since other signal-dependent noise sources only have very limited impact (less than 3\% according to \cite{gow2007comprehensive, janesick1987scientific}).
    Since the incident photon numbers follow the Poisson distribution strictly as described in Eq.~\eqref{eq:shot_noise}, we can accurately synthesize the signal-dependent noise as
    \begin{align}
        &Y = K_d K_a I, \\
        &(I + \hat{N_p}) = \mathcal{P}\left(\frac{Y}{K_d K_a}\right),
        \label{eq:shot_generation}
    \end{align}
    where $Y$ is the clean image obtained by removing all noise in Eq.~\eqref{eq:decomposed_overview}, and $\hat{N_p}$ is the sampled signal-dependent noise (photon shot noise).
    To synthesize the photon shot noise, we invert the digital numbers $Y$ to the photon numbers by dividing the total gain $K_d K_a$. Then, we sample the signal-dependent noise from the Poisson distribution.
    The total gain $K_d K_a$ can be estimated by the linear curve between the mean and variance of flat-field frames \cite{photonTransfer}.
    \paragraph{Signal-independent noise.} For the signal-independent noise, we capture dark frames in the dark room to construct the dark-frame database, and the synthetic noise is sampled directly from the dark-frame database.
    In each shooting of the dark frame, all signal-dependent noise naturally disappears as we block all incident light. Formally, it can be formulated by removing the signal-dependent component in Eq.~\eqref{eq:decomposed_overview}:
    \begin{align}
        &B = K_d K_a N_1 + K_d N_2 + K_d N_q, \\
        &\hat{N}_\mathrm{independent} \leftarrow \text{RandomSampling}(B),
    \end{align}
    where $B$ is the dark frame, and $\hat{N}_\mathrm{independent}$ is the synthetic signal-independent noise.
    All actual signal-independent noise is accumulated and saved in the shooting of the dark frames. As a result, the synthetic signal-independent noise is most accurate and contains all kinds of noise types that cannot be modeled explicitly before.

    Moreover, it is convenient and feasible to sample actual signal-independent noise from dark-frame databases of different devices with various exposure and ISO settings.
    For a popular camera sensor, each raw image contains more than 1M pixels, which indicates more than 1M noise values are sampled from the actual noise distribution at each shoot. We can easily obtain sufficient signal-independent noise samples to form the signal-dependent noise database.
 
    By synthesizing the signal-dependent and -independent components separately, for a given clean image $Y$, we can generate the synthetic noisy image via Eq.~\eqref{eq:decomposed_overview}.
    \begin{equation}
        \hat{D} = K_d K_a (I + \hat{N_p}) + \hat{N}_{independent},
        \label{eq:synthesis_overview}
    \end{equation}
    where $\hat{D}$ is the synthetic noisy image.

    \renewcommand \arraystretch{1.2}
\begin{table}[t]
    \centering
    \caption{The comparison between Poisson-Gaussian (P-G) noise model, ELD noise model, pixel-wise sampling, patch sampling, and pattern-aligned patch sampling (PAP) on the ELD \cite{eld} SonyA7S2 dataset. ``LB'' and ``HB'' denote low-bit (14-bit) and high-bit (32-bit) respectively. See Section~\ref{sec:implementaion} for more implementation details.}
    \label{tb:naive_sampling}
    \small
    \begin{tabular}{lccc}
        \toprule
        ~ & $\times 1$ & $\times 100$ & $\times 200$ \\
        Model & PSNR / SSIM & PSNR / SSIM & PSNR / SSIM \\
        \midrule
        P-G \cite{FoiPG} & 53.79 / 0.997 & 46.96 / 0.926 & 43.84 / 0.856 \\
        P-G + LB & 53.59 / 0.997 & 46.35 / 0.944 & 42.98 / 0.872 \\
        ELD \cite{eld} & 53.43 / 0.997 & 47.77 / 0.970 & 45.46 / 0.940 \\
        ELD + LB & 53.41 / 0.997 & 45.74 / 0.920 & 42.35 / 0.847 \\\midrule \midrule
        Pixel-wise & 53.90 / 0.997 & 45.71 / 0.917 & 42.57 / 0.846 \\
        Patch & 53.93 / 0.997 & 45.83 / 0.918 & 42.70 / 0.846 \\
        PAP & 53.91 / 0.997 & 47.13 / 0.958 & 44.36 / 0.906 \\
        PAP + HB & 53.92 / 0.997 & 47.90 / 0.973 & 45.47 / 0.940 \\
        \bottomrule
    \end{tabular}
\end{table}
\renewcommand \arraystretch{1.}


    \subsection{Treating signal-independent noise separately does not work well}
    We first try a naive implementation by treating each dark-frame pixel value as an individual signal-independent noise sample and save them separately in the noise database. We use SonyA7S2 for this experiment and evaluate on the ELD dataset.
    
    Specifically, for the signal-dependent noise, we synthesize it as described in Eq.~\eqref{eq:shot_generation}.
    For synthesizing the signal-independent noise, 10 dark frames are captured under no-light condition for each ISO setting with the default exposure time to construct the dark-frame database. The noise values are saved equally and signal-independent noise is synthesized by sampling pixels randomly from the dark-frame database.
    The results are shown in Table \ref{tb:naive_sampling}.
    Benefited from the noise sampling from the actual noise database, even the naive implementation outperforms the state-of-the-art ELD noise model \cite{eld} in normal exposure ratio (ratio $\times 1$), which is the exposure ratio we commonly used in daily life. But for more challenging extreme low-light environments (ratio $\times 100$, $\times 200$), we found directly pixel-wise sampling for the signal-independent noise cannot match the performances of statistical methods. We found that it is due to two issues:
    \begin{enumerate}
        \item The above naive signal-independent noise sampling destroys their spatially-correlated characteristics (\eg, row noise, fix pattern noise), which become significant in low-light conditions \cite{eld, enhancingNJU}.
 
        \item The signal-independent noise is the dominant noise in low-light conditions, but the dark frames have been quantized into lower bits (10- or 14-bit images), which destroy the natural signal-independent noise distributions significantly.
    \end{enumerate}
    \subsection{Pattern-aligned patch sampling and high-bit reconstruction}
    To tackle the two problems in naive signal-independent noise sampling, we propose two techniques to better synthesize such noise: pattern-aligned patch sampling and high-bit reconstruction.
    \paragraph{pattern-aligned patch sampling.}
    The spatially-correlated noise becomes significant under low-light conditions.
    To preserve the spatially-correlated noise, we use patch sampling, which synthesizes the signal-independent noise by cropping patches from the constructed dark-frame database. In this way, the spatially-correlated noise is preserved in the dark-frame patches.
    Moreover, we observe that the raw Bayer pattern also affects the noise distribution, we further propose pattern-aligned patch sampling (PAP), which ensure that dark-frame patches have the exactly aligned Bayer pattern as the clean image to be corrupted. 
    Table~\ref{tb:naive_sampling} shows the results of using pixel-wise sampling, patch sampling and pattern-aligned patch sampling (PAP) for synthesizing signal-independent noise.
    The proposed pattern-aligned patch sampling helps to improve the denoising performance under the extreme low-light condition.

    \paragraph{The effect of high-bit synthetic noise.}
    In extreme low-light conditions, the signal-independent noise becomes the dominant noise, which requires accurate and high-bit synthetic noise. To show that, in Table \ref{tb:naive_sampling}, we quantize the signal-independent component of the physics-based methods to 14-bit (the same as the dark frames), which is denoted as ``LB''. The physics-based methods show an obvious performance drop when using lower-bit synthetic noise, especially for low-light conditions (\eg, $\times 100, \times 200$). Therefore, high-bit signal-independent noise plays a crucial role in noise synthesis for low-light environments.

    \paragraph{High-bit reconstruction.}
    Theoretically, we can dump dark frames before the quantization operation in the sensor processing pipeline or save high-bit dark frames (\eg 32-bit) to preserve actual signal-independent noise. However, most CMOS sensor manufacturers do not allow researchers to have access to such raw data.
    Therefore, we propose a high-bit reconstruction method to inverse the quantization first and synthesize high-bit values to refine the noise patches sampled by the pattern-aligned patch sampling.

\begin{figure}[t]
        \centering
        \begin{subfigure}[b]{0.22\textwidth}
            \includegraphics[width=\textwidth]{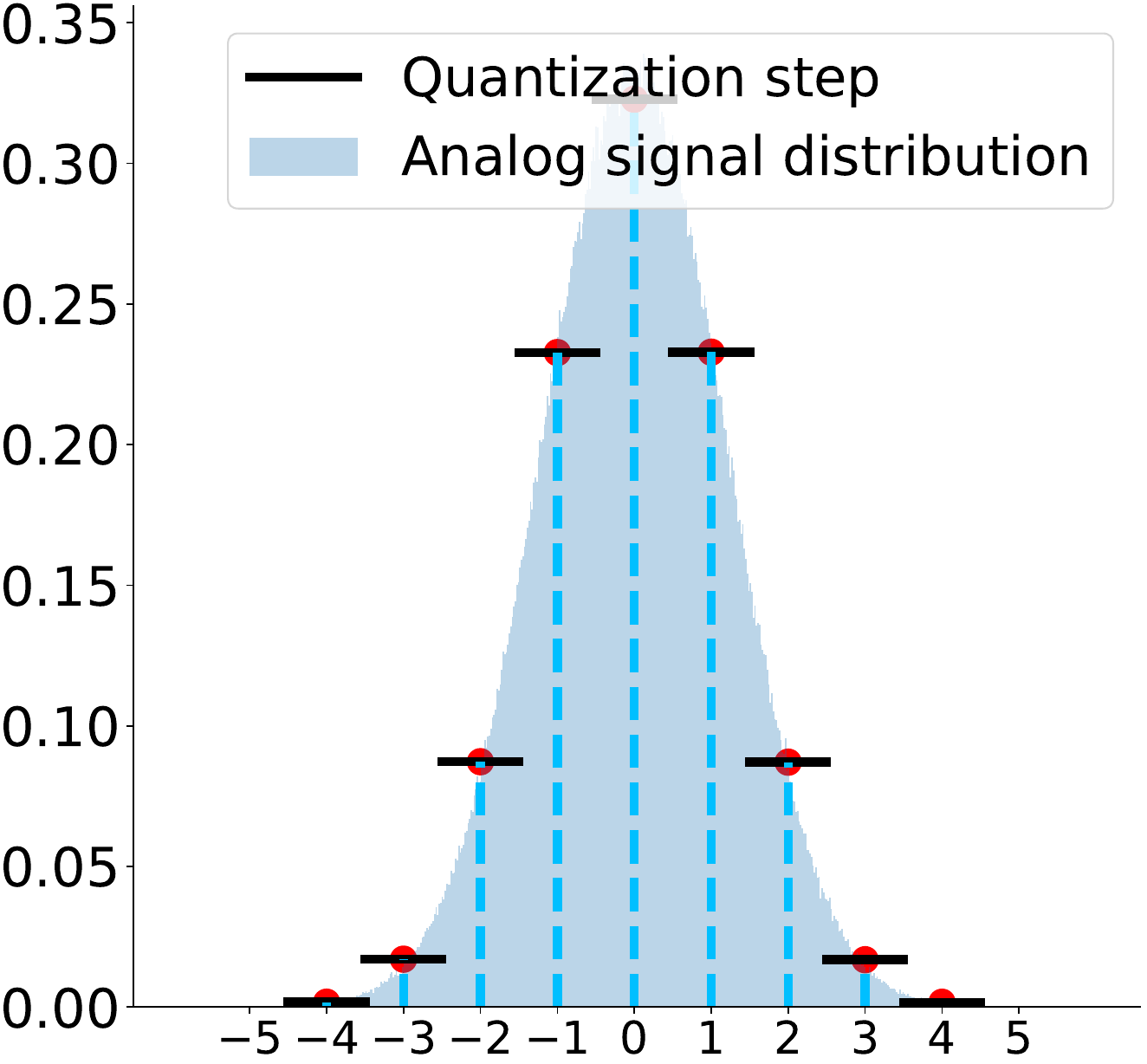}
            \caption{Signal quantization.}
            \label{fig:toy_quant}
        \end{subfigure}
        ~~
        \begin{subfigure}[b]{0.22\textwidth}
            \includegraphics[width=\textwidth]{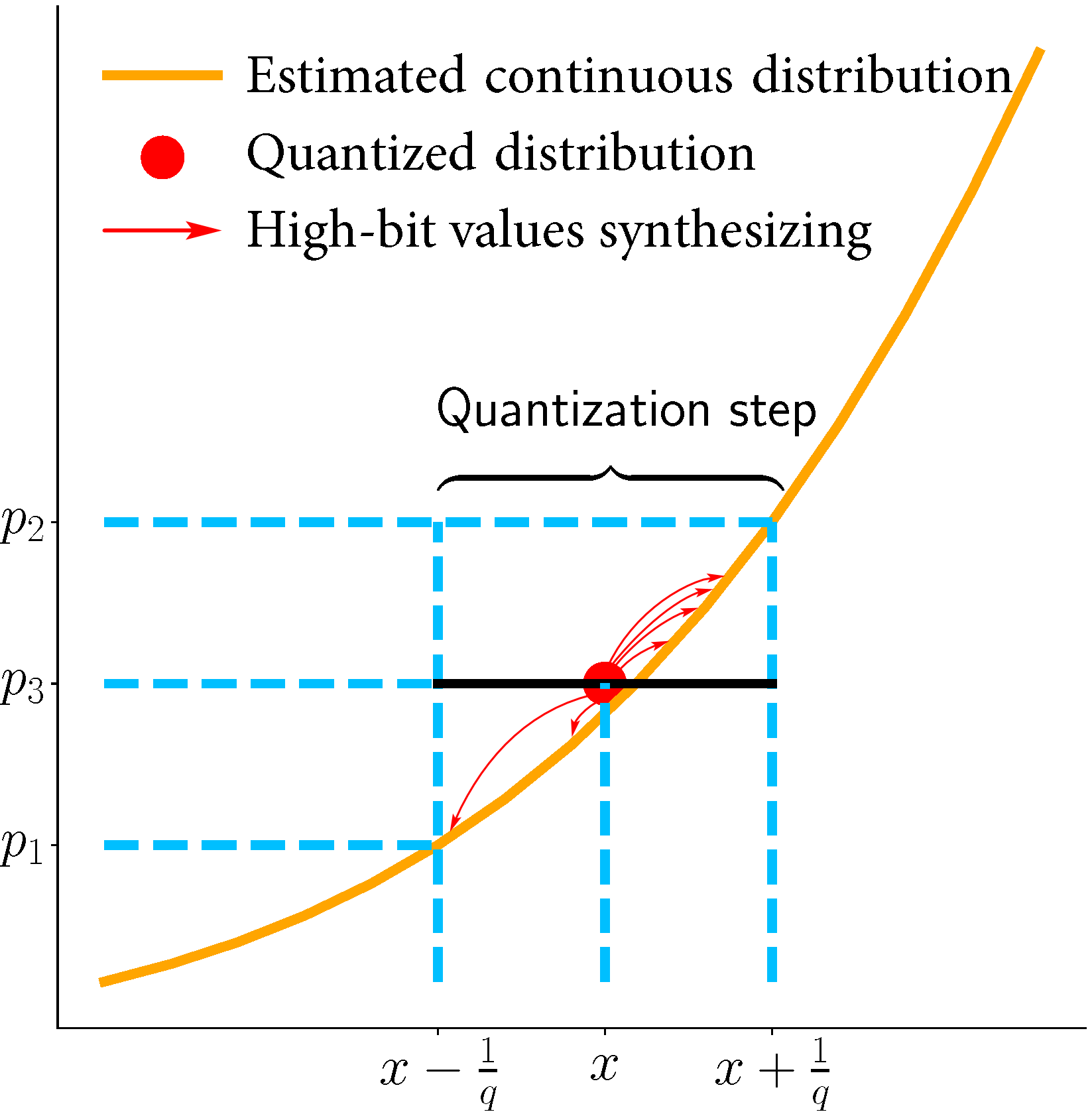}
            \caption{High-bit reconstruction.}
            \label{fig:continous_rec}
        \end{subfigure}
        \caption{The toy example of signal quantization and high-bit reconstruction for the quantized value $x$ within the quantization step.}
        \label{fig:pattern_crop_continous}
    \end{figure}

    Figure~\ref{fig:toy_quant} shows a toy example illustrating the quantization effect for the continuous analog signal distribution of camera sensors. Figure~\ref{fig:continous_rec} shows the details within each quantization step.
    The quantization process maps a set of continuous values in $[x - \frac{1}{q}, x + \frac{1}{q}]$ to a single value $x$, which makes our collected dark frames have lower bits.
    To reconstruct the high-bit dark frames, we first estimate the continuous distribution of the low-bit noise sampled from the dark frames. 
    Then, as shown in Figure~\ref{fig:continous_rec}, according to the estimated continuous distribution, we sample high-bit value (32-bit) within the quantization step $[x - \frac{1}{q}, x + \frac{1}{q}]$ for each low-bit discrete value $x$, and use the sampled high-bit values to replace the $x$ in the dark frames. The sample probability of high-bit values is obtained by normalizing the estimated continuous distribution within the quantization step.
    In this way, the dark frames are reconstructed to higher bits while keeping the characteristics of the real noise.

    We use statistical models to estimate the continuous distribution of quantized noise in the dark-frame database. Instead of modeling it with a single distribution (\eg, Gaussian \cite{FoiPG}, Tukey Lambda \cite{eld}), we construct a continuous distribution set which contains 5 bell-shaped distributions, including Student's t, Weibull, Tukey lambda, Gaussian, and Gamma distribution, for better modeling noise of different sensors.
    We use the distribution set to fit the noise in the dark frames and identify the distribution that can best describe the noise distribution.

    \skipspace
    \section{Experiments}

    We systematically compare existing DNN-based methods, physics-based statistical methods, and our method for noise synthesis on SIDD mobile dataset \cite{sidd} and ELD DSLR dataset \cite{eld}, and use the synthesized images for training denoising networks.
    We also demonstrate the effectiveness and generalization of our method over different camera sensors and lighting conditions.
    \subsection{Implementation details}\label{sec:implementaion}
    
    \paragraph{Evaluation datasets.} We follow previous works \cite{eld, noiseflow,CameraNoiseModeling} and select two representative datasets to systematically evaluate our methods and compare with existing DNN-based and physics-based methods.

    \begin{enumerate}
        \item SIDD \cite{sidd} is a mobile dataset, which is collected by five different smartphones under three lighting conditions (low-light, normal and high exposure). We use three recent smartphones (iPhone 7, Google Pixel, Samsung Galaxy S6 Edge) in our experiments. Most recent DNN-based methods \cite{noiseflow, CameraNoiseModeling, dualNoiseGan20} have demonstrated their effectiveness on SIDD dataset.
        \item  ELD \cite{eld} is a recent DSLR dataset, which is captured over a wide range exposure ratio ($\times 1 \sim \times 200$). We evaluate our method on two popular cameras (SonyA7S2 and NikonD850) over all exposure ratios.
    \end{enumerate}
    \renewcommand{\arraystretch}{1.2} 
\begin{table*}[!t]
    \centering
    \caption{Quantitative results (PSNR/SSIM) of DNN-based, physics-based, and our real-noise-based methods on SIDD dataset. We use different methods to synthesize noise during the training of denoising models. All other settings remain the same. ``P-G'', ``Pixel-wise'', and ``PAP + HB'' represent Poisson-Gaussian noise model, pixel-wise sampling, and pattern-aligned patch sampling with high-bit refinement respectively. The \ctr{red} color indicates the best result and the \ctb{blue} color indicates the second best result.}

    \small
    \setlength{\tabcolsep}{1mm}{  
    \begin{tabular}{C{.08\linewidth}| C{.05\linewidth}|C{.09\linewidth}|C{.11\linewidth}|C{.1\linewidth}|C{.1\linewidth}|C{.075\linewidth}|C{.08\linewidth}|C{.08\linewidth}|C{.08\linewidth}}
        \hline
        \xrowht{10pt}
        \multirow{3}{*}{Camera} & \multirow{3}{*}{Index} & \multicolumn{2}{c|}{DNN-based} & \multicolumn{3}{c|}{Physics-based}  & \multicolumn{3}{c}{Real-noise-based} \\ \cline{3-10}
        & & Noise FLow \newline \cite{noiseflow}  & Camera-Aware \newline  GAN \cite{CameraNoiseModeling}& In-Camera \newline P-G & Calibrated \newline P-G & \multirow{2}{*}{ELD \cite{eld}}  & \multirow{2}{*}{Paired data} & \multirow{2}{*}{Pixel-wise}& \multirow{2}{*}{PAP + HB}\\ \hline
        \multirow{2}{*}{S6}
        &PSNR&44.68&45.00&34.64&45.38&\ctb{45.39}&44.35&45.33&\ctr{45.62}\\\cline{2-10}
        &SSIM&0.973&0.975&0.720&0.976&\ctb{0.977}&0.971&0.976&\ctr{0.978}\\\hline
        \multirow{2}{*}{IP}
        &PSNR&48.29&55.48&44.58&56.71&56.73&55.64&\ctb{56.76}&\ctr{56.79}\\\cline{2-10}
        &SSIM&0.983&0.994&0.960&0.996&0.996&0.995&\ctb{0.996}&\ctr{0.996}\\\hline
        \multirow{2}{*}{GP}
        &PSNR&45.15&45.96&36.65&46.48&46.43&44.58&\ctb{46.52}&\ctr{46.56}\\\cline{2-10}
        &SSIM&0.962&0.975&0.776&0.976&0.977&0.955&\ctb{0.977}&\ctr{0.978}\\\hline
    \end{tabular}}
    \label{tb:sidd}
    \vspace{-10pt}
\end{table*}
    \paragraph{Compared methods.}
    \begin{enumerate}
        \item Poisson-Gaussian (P-G) distribution \cite{FoiPG} is the most typical noise model for raw image denoising. We use the accurate P-G noise model for comparison instead of approximating the Poisson distribution to the Gaussian distribution.
        
        \item ELD noise model \cite{eld} is the recently developed method for low-light environments. We have reproduced the results in ELD with the authors' help and modified several lines of code to synthesize noise for a wider ratio range ($\times 1 \sim \times 200$).
        
        \item Representative DNN-based methods including both recent GAN-based \cite{CameraNoiseModeling} and flow-based \cite{noiseflow} methods are used for comparison. We use their released models to synthesize noise on SIDD dataset and train a new model for ELD dataset by adopting the authors' code.
    \end{enumerate}
    \paragraph{Training and testing.}
    In all experiments, we adopt the U-shape net in SID \cite{sid} as our backbone for training denoising models with synthesized noisy images.
    For both SIDD and ELD dataset, we pack raw images to four channels and randomly sample non-overlapped $512 \times 512$ patches with rotation and flipping augmentation. We follow the common training setting: $L_1$ loss function, Adam optimizer with default hyper-parameters, training from scratch, batch size 1, and initial learning rate of $10^{-4}$. The models are trained for 30k iterations and the learning rate halves twice at iterations 15k and 25k.
    During the training, we sample the ISO uniformly from the set $\{2^n 100 | n \in \mathbb{N}, 0 \leq n \leq 5\}$. For each ISO, we capture 10 dark frames to construct the dark-frame database and use the proposed pattern-aligned patch sampling with high-bit reconstruction to synthesize realistic signal-independent noise. The synthesis of the signal-dependent noise follows Eq. \eqref{eq:shot_generation}.

    For the SIDD dataset, three scenes (02, 03 and 05) are kept for testing. For evaluating on ELD dataset, we train models on SID Sony dataset and use the official testing code for all cameras and exposure ratios.

    \renewcommand \arraystretch{1.2}
\begin{table*}[t]
    \centering
    \caption{Quantitative results (PSNR/SSIM) of different methods on ELD dataset with four different exposure ratios. All settings are kept the same except the method for synthesizing noise to create training data for denoising model training. ``P-G'', ``Pixel-wise'', and ``PAP + HB'' represent Poisson-Gaussian noise model, pixel-wise sampling, and our proposed pattern-aligned patch sampling with high-bit reconstruction, respectively.}
    \label{tb:sid}
    \small
    \setlength{\tabcolsep}{1mm}{
    \begin{tabular}{C{.09\linewidth}|C{.08\linewidth}|C{.05\linewidth}|C{.11\linewidth}|C{.09\linewidth}|C{.09\linewidth}|C{.11\linewidth}|C{.085\linewidth}|C{.085\linewidth}}
        \hline
        \multirow{2}{*}{Camera} & \multirow{2}{*}{Ratio} & \multirow{2}{*}{Index}  &   \multicolumn{1}{c|}{DNN-based} & \multicolumn{2}{c|}{Physics-based} & \multicolumn{3}{c}{Real-noise-based} \\ \cline{4-9}
        & & & Noise Flow \cite{noiseflow} & P-G \cite{FoiPG} & ELD \cite{eld}& Paired data \cite{sid} & Pixel-wise & PAP + HB \\\hline
        \multirow{8}{*}{SonyA7S2}
        & \multirow{2}{*}{$\times 1$}
          &PSNR&41.20&53.79&53.43&NA&\ctb{53.90}&\ctr{53.92}\\\cline{3-9}
        & &SSIM&0.842&0.997&0.997&NA&\ctb{0.997}&\ctr{0.997}\\\cline{2-9}
        ~ & \multirow{2}{*}{$\times 10$}
          &PSNR&40.79&\ctb{51.79}&51.70&NA&51.73&\ctr{52.14}\\\cline{3-9}
        & &SSIM&0.836&\ctb{0.992}&0.993&NA&0.992&\ctr{0.994}\\\cline{2-9}
        ~ & \multirow{2}{*}{$\times 100$}
          &PSNR&38.68&46.96&\ctb{47.77}&46.73&45.71&\ctr{47.90}\\\cline{3-9}
        & &SSIM&0.793&0.926&\ctb{0.970}&0.970&0.917&\ctr{0.973}\\\cline{2-9}
        ~ & \multirow{2}{*}{$\times 200$}
          &PSNR&36.30&42.84&\ctb{45.46}&44.57&42.57&\ctr{45.47}\\\cline{3-9}
        & &SSIM&0.713&0.856&\ctb{0.940}&0.935&0.846&\ctr{0.940}\\\hline
        \multirow{8}{*}{NikonD850}
        ~ & \multirow{2}{*}{$\times 1$}
          &PSNR&39.20&\ctb{48.84}&48.60&NA&48.57&\ctr{49.15}\\\cline{3-9}
        & &SSIM&0.831&\ctb{0.990}&0.989&NA&0.990&\ctr{0.991}\\\cline{2-9}
        ~ & \multirow{2}{*}{$\times 10$}
          &PSNR&37.97&\ctb{47.01}&46.81&NA&46.87&\ctr{47.34}\\\cline{3-9}
        & &SSIM&0.806&\ctb{0.979}&0.979&NA&0.981&\ctr{0.984}\\\cline{2-9}
        ~ & \multirow{2}{*}{$\times 100$}
          &PSNR&32.87&41.61&\ctb{42.08}&41.13&41.83&\ctr{42.56}\\\cline{3-9}
        & &SSIM&0.694&0.877&\ctb{0.910}&0.930&0.920&\ctr{0.925}\\\cline{2-9}
        ~ & \multirow{2}{*}{$\times 200$}
          &PSNR&30.40&39.19&\ctb{39.99}&39.36&39.43&\ctr{40.35}\\\cline{3-9}
        & &SSIM&0.583&0.825&\ctb{0.877}&0.909&0.879&\ctr{0.889}\\\hline

    \end{tabular}}
    \vspace{-10pt}
\end{table*}

    \begin{figure}[t]
        \begin{center}
            \includegraphics[width=\linewidth]{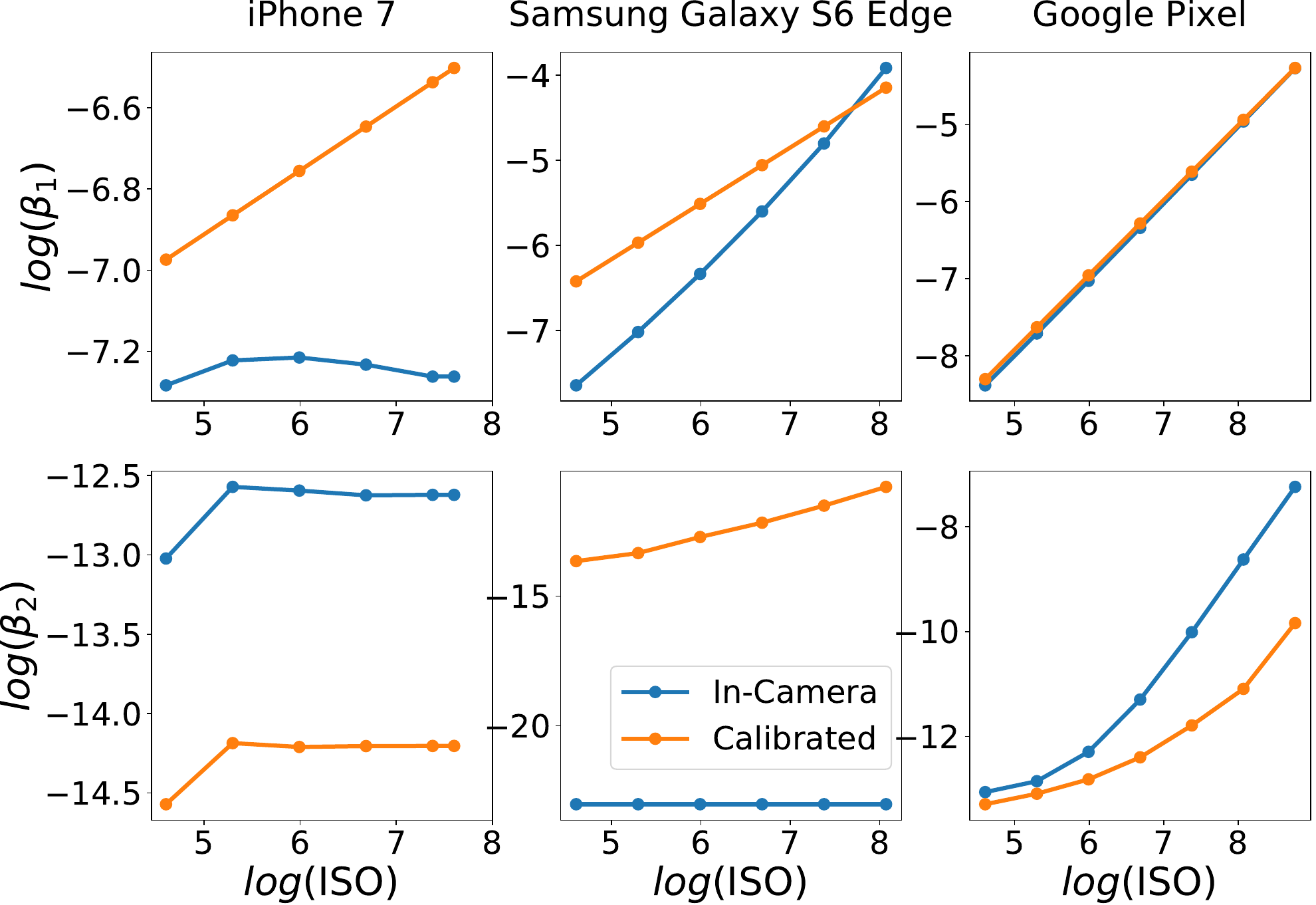}
        \end{center}
        \vspace{-10pt}
        \caption{The comparison of in-camera and calibrated noise profiles on SIDD dataset. Both ISO and noise profile ($\beta_1$, $\beta_2$) are transformed to the log domain for better visualization. The original ISOs are sampled from the interval [100, 6400] (the corresponding $log(\text{ISO}) \in$ [4.61, 8.76]).}
        \label{fig:sidd_calibration}
    \end{figure}
    \subsection{SIDD dataset}
    \paragraph{Parameters calibration.}
    The in-camera noise profile for Poisson-Gaussian noise model can be obtained directly from the meta information of DNG file \cite{dng}. They have been used widely in existing works.
    However, we found that the in-camera noise profiles are not calibrated seriously.

    We collect the three recent smartphones used in SIDD dataset and estimate parameters for the P-G noise model as described in the original paper \cite{FoiPG}.
    As shown in Figure~\ref{fig:sidd_calibration}, except the parameters $\beta1$ of Google Pixel, all other in-camera noise parameters show large differences compared with our calibrated parameters. In Table~\ref{tb:sidd}, we use the in-camera and calibrated noise parameters to synthesize noise for denoising model training and keep all other settings the same. The results show that using the calibrated parameters performs much better than using the in-camera parameters. This indicates that the in-camera noise parameters are not accurate.

    We also calibrate the noise parameters for the ELD noise models. It uses the Tukey Lambda (TL) distribution family \cite{joiner1971some} to fit the read noise distribution and use the Probability Plot Correlation Coefficient (PPCC) \cite{ppcc1975} test to find the best shape parameter.
    We found that, for the mobile sensors, the estimated shape parameter is quite close to 0.14, which is an approximate Gaussian distribution in the TL distribution family. 
    As a result, ELD noise model reduces to the P-G noise model with additional row noise and quantization noise.
    \paragraph{Denoising results.}
    We train a denoising model for each noise synthesis method with the same training setting. The denoising results on SIDD dataset are summarized in Table~\ref{tb:sidd}.

    Same as reported in previous works, the in-camera P-G noise mode cannot outperform DNN-based methods, especially for GAN-based models. However, unluckily, all previous works are built on inaccurate noise profiles, since the re-calibrated P-G noise model shows large PSNR improvements on the SIDD dataset. It even works much better than all recent DNN-based methods in our experiments.
    As described before, the TL distribution in the ELD noise model reduces to the Gaussian distribution. Therefore, the results of the ELD noise model on all three cameras are quite similar to the P-G noise model.

    Our method consistently outperforms both DNN-based and statistical methods over all three mobile cameras. Since the SIDD dataset is captured under the normal exposure ratio ($\times 1$), the naive pixel-wise sampling can produce comparable results as the well-calibrated P-G noise models. And , using the high-bit reconstruction also shows consistent improvements.

    \begin{figure*}[t]	
      \begin{center}	
          \includegraphics[width=\linewidth]{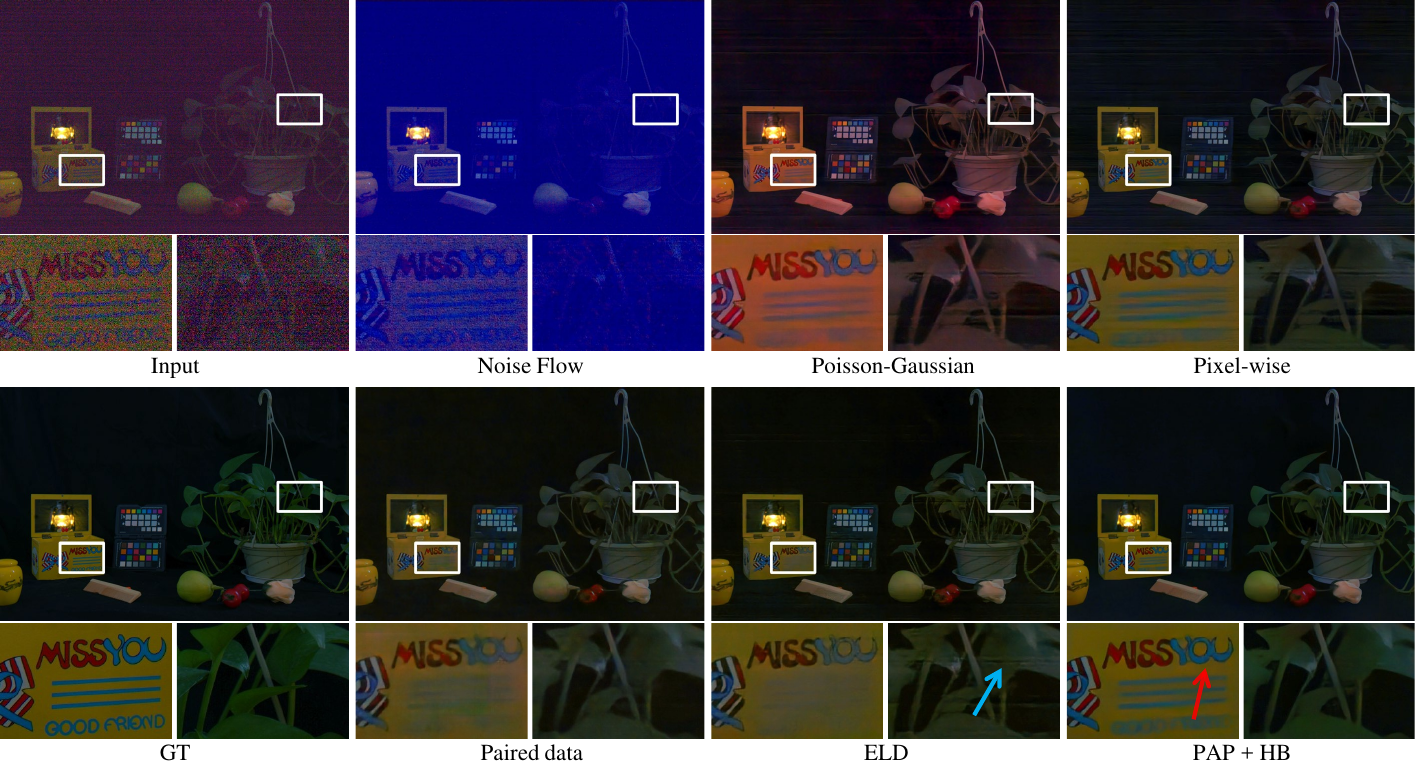}	
      \end{center}	
      \vspace{-10pt}	
      \caption{The qualitative results on ELD dataset. We use different methods to synthesize the noise for creating training images. All other settings are kept the same when training the denoising models. 
      Our method helps the denoising model preserve more textures and vivid colors (pointed by the red arrow). Moreover, only our method (PAP + HB) and the paired data \cite{sid} can remove the fixed pattern noise successfully (pointed by the blue arrow), since the physics-based methods are unable to synthesize the unmodeled noise. See the text for more analysis.}
      \label{fig:sid_vis}
  \end{figure*}

    \subsection{ELD dataset}
    \paragraph{Parameters calibration.}
    The noise profiles are not provided in both SonyA7S2 and NikonD850 cameras. Similar to the experiment on SIDD dataset, we use the same model of cameras and calibrate noise profiles for statistical methods (P-G and ELD noise models).

    For DNN-based methods, we use Noise Flow \cite{noiseflow} for comparison and modify the authors' code for training the denoising model on the SID dataset. We found that the training process of Noise Flow on the extreme low-light dataset is quite unstable. See Section~\ref{sec:implementaion} for more implementation details.
    \paragraph{Denoising results.}
    Table~\ref{tb:sid} shows the denoising performances over four exposure ratios of the denoising model trained with data from different noise synthesis methods.
    The DNN-based method shows poor generalization on the extreme low-light datasets. It suffers from severe color distortion as shown in Figure~\ref{fig:sid_vis}. 
    
    For the physics-based methods, the ELD noise model outperforms the P-G noise model in extreme low-light environments ($\times 100, \times 200$). But for the normal lighting conditions ($\times 1, \times 10$), the P-G noise model shows better performances. The results for exposure ratio 1 and 10 are not available (NA) since there are no corresponding training pairs for ratio 1 and 10 in the SID dataset. 
    
   Our methods (denoted by PAP + HB) consistently outperforms all existing noise synthesis methods over different cameras and exposure ratios. Table~\ref{tb:sid} shows that the high-bit reconstruction plays an important role in our method, especially in extreme low-light environments. Some qualitative results are shown in Figure~\ref{fig:sid_vis}. The denoising model trained with our synthetic noise can recover more textures and vivid colors compared with existing methods.
   
   Since our method can synthesize all kinds of signal-independent noise, it helps the denoising model remove the fixed pattern noise successfully even though we do not model it explicitly. The model trained with the real noise (paired data) can also remove it. But physics-based methods can only synthesize the noise that they are able to model. As a result, the denoising model trained with them cannot deal with the unmodeled noise. 

    \section{Conclusions}
    
    In this paper, we propose a new perspective for raw image noise synthesis. The synthetic noise from the presented method is accurate inherently since it benefits from the actual noise. Two techniques, pattern-aligned patch sampling and high-bit reconstruction, for the signal-independent noise enhance the synthesis performance and generalization in the low-light environments.
    Our method is both generic and efficient. On both SIDD and ELD datasets, our method outperforms recent DNN-based methods and physics-based methods over different cameras and lighting conditions.
    Moreover, through systematic comparisons, we found existing comparisons of DNN-based methods and physics-based methods are built on inaccurate noise parameters. After careful noise parameters calibration, DNN-based methods surprisingly underperform the physics-based methods with a clear gap, which is opposite to the recent advances.
    
    \section{Acknowledgement}
    This work is supported in part by the General Research Fund through the Research Grants Council of Hong Kong under Grants (Nos. 14204021, 14208417, 14207319, 14202217, 14203118, 14208619, ), in part by Research Impact Fund Grant No. R5001-18, in part by CUHK Strategic Fund.

    {\small
    \bibliographystyle{ieee_fullname}
    \bibliography{egbib}
    }

\onecolumn
\section{Details for noise parameters calibration}
Here, we provide more details on noise parameters calibration for three smartphone cameras in the SIDD dataset. Following the DNG document \cite{dng}, we estimate two noise parameters $(\beta_1, \beta_2)$ for each sensor under the typical Poisson-Gaussian distribution. Specifically, $\beta_1$ denotes the total gain factor inner the camera and $\beta_2$ is the variance (Gaussian distribution) for the signal independent noise. For $\beta_1$, we first capture a series of \textit{flat-field frames} by adjusting the uniform illumination and use the Photo Transfer method \cite{photonTransfer} to fit the linear relationship between the mean and variance of the \textit{flat-field frames}. For $\beta_2$, we capture \textit{black frames} for each ISO settings and remove their black levels. The variance parameter $\beta_2$ can be estimated under the Gaussian distribution.

\section{More qualitative results.}
\begin{figure*}[t]
    \centering
    \includegraphics[width=0.8\textwidth]{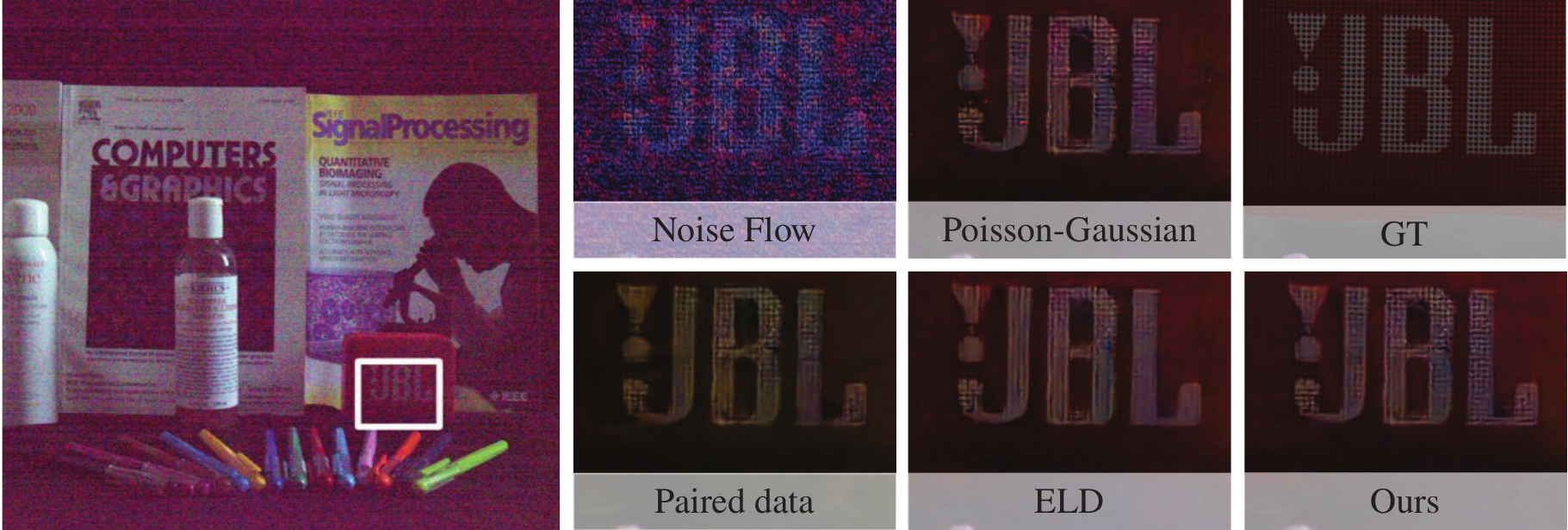}
    \includegraphics[width=0.8\textwidth]{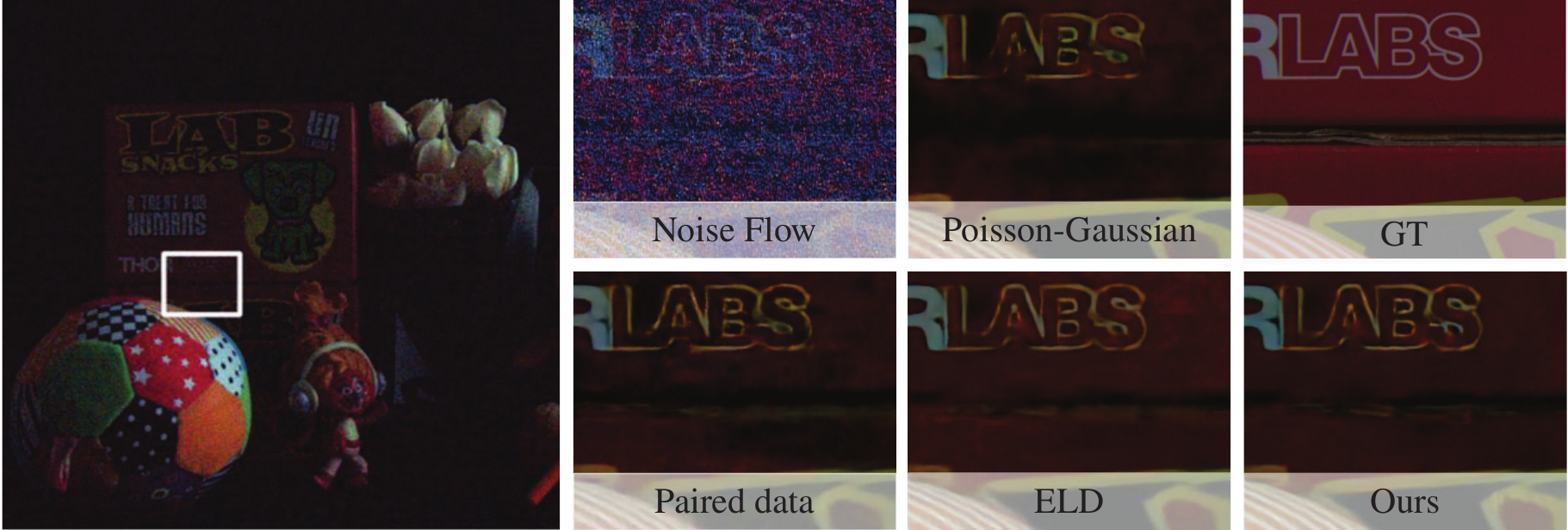}
  
    \caption{Qualitative results from the ELD dataset (\textbf{top}: scene 6, \textbf{bottom}: scene 1).}
    \label{fig:sets}
\end{figure*}

In fig.~\ref{fig:sets}, we show more qualitative results from the ELD dataset. Our method can produce more textures than other methods. And, the color around the dark areas also can be preserved well.

\end{document}